\documentclass[a4paper,fleqn,usenatbib]{mnras}

\usepackage{newtxtext,newtxmath}
\usepackage[T1]{fontenc}
\usepackage{ae,aecompl}

\usepackage{graphicx}	% Including figure files
\usepackage{amsmath}	% Advanced maths commands
\usepackage{amssymb}	% Extra maths symbols
\usepackage{threeparttable}% Three-part table with table notes
\usepackage{adjustbox}
\usepackage{color}
\usepackage{pdflscape}

\newcommand{\hii}{H\,{\sc ii}}
\newcommand{\bra}{Br\,$\alpha$} 
\newcommand{\mbra}{{\rm Br}\,\alpha}

\newcommand{\msun}{$\rm M_\odot$}
\newcommand{\kms}{$\rm km\,s^{-1}$}

\title[Ionized Gas Motions in the NGC~253 Starburst]{Unveiling Kinematic Structure in the Starburst Heart of NGC 253}

\author[D. P. Cohen et al.]{
Daniel P. Cohen,$^{1}$\thanks{E-mail: dcohen@astro.ucla.edu}
Jean  L. Turner,$^{1}$
S. Michelle Consiglio$^{1}$
\\
$^{1}$University of California, Los Angeles, CA 90095-1547, USA 
}

\date{Accepted 2020 January 24. Received 2020 Jan 15; in original form 2019 December 2}

\pubyear{2020}

\begin{document}
\label{firstpage}
\pagerange{\pageref{firstpage}--\pageref{lastpage}}
\maketitle

\begin{abstract}
We investigate the kinematics of
ionized gas within the
nuclear starburst of NGC~253
with observations of the Brackett $\alpha$ recombination line at 4.05 $\mu$m. The goal is to distinguish motions driven by star-formation feedback from gravitational motions induced by the central mass structure.
Using NIRSPEC on Keck II, we obtained 30 spectra through a $0\farcs5$ slit stepped across the central $\sim$5\arcsec$\times$25{\arcsec} (85 $\times$ 425 pc) region to produce a spectral cube.
The Br$\alpha$ emission resolves into four nuclear sources: 
S1 at the infrared core (IRC), N1 at the radio core near nonthermal source TH2, and the fainter sources N2 and N3 in the northeast. 
The line profile is characterized by a primary component with $\Delta v_{\mathrm{primary}}$$\sim$90-130 \kms\ (FWHM) on top of a broad {blue} wing with $\Delta v_{\mathrm{broad}}$$\sim$300-350 \kms, and an additional redshifted narrow component in the west. The velocity field generated from our cube reveals several distinct patterns. A mean NE-SW velocity gradient of +10 \kms\ arcsec$^{-1}$ along the major axis traces the solid-body rotation curve of the nuclear disk. At the radio core, isovelocity contours become S-shaped, indicating the presence of secondary nuclear bar of total extent $\sim$5\arcsec (90 pc). 
The symmetry of the bar places the galactic center near the radio peak TH2 of the galaxy rather than the IRC, and makes this the most likely location of a SMBH. A third kinematic substructure is formed by blueshifted gas on the southeast side of the IRC. This feature likely traces a $\sim$100-250 {\kms} starburst-driven outflow, linking the IRC to the galactic wind observed on kpc scales. 
\end{abstract}


%%%%%%%%%%%%%%%%%%%%%%%%%%%%%%%%%%%%%%%%%%%%%%%%%%

%%%%%%%%%%%%%%%%% BODY OF PAPER %%%%%%%%%%%%%%%%%%

\section{INTRODUCTION} 

The nuclear regions of disk galaxies are singular environments, in the sense that they serve as a distinct destination for the
inward flow of matter in disk galaxies.
Internal secular processes shape the 
continuing evolution of present-day disk galaxies \citep[][]{kormendy2004}.
Radial gas flow induced by nonaxisymmetric structures such as bars, together with star formation 
and feedback from active galactic nuclei, 
all serve to shape a constantly evolving central mass structure in spiral galaxies.

Due to the inward flow of gas, barred galaxies often host intense nuclear star formation or 
starbursts, 
often taking the form of nuclear rings of super star clusters  
\citep[SSCs; e.g.,][]{athana1984,buta1996a,buta1999,boker2008,comeron2010}, as predicted by simulations
 \citep{athana1992,shlosman2002a,regan2003,li2015,sormani2018,seo2019}.
Star formation in barred galaxies can serve as a sink of the inwardly drifting gas, thus preventing
this gas  from reaching  a central supermassive black hole. 
However, star formation is never a pure gas sink; it is inefficient and can disperse gas via winds. Young 
SSCs are likely responsible for the multi-phase, large-scale galactic winds observed in 
starburst galaxies \citep[e.g.,][]{heckman2001}. Hydrodynamic simulations of barred galaxies find that such intense feedback can shape the growth of galactic bulges \citep[e.g.,][]{renaud2013,athana2013,carles2016,li2015,seo2019}.
AGN feedback 
 from an accreting SMBH {can also} regulate star formation in the galactic center and its host bulge \citep[e.g.,][]{robichaud2017}. 
On the other hand, remnant, post star-forming molecular clouds can potentially assist the inward
drift of their natal embedded clusters via a nuclear bar \citep{tsai2013}. These are only a few of the possible effects governing gas flows and secular evolution in the centers of disk galaxies.

To investigate the processes of gas inflow and feedback in a barred galaxy, 
we have studied ionized gas kinematics 
in NGC~253, one of the closest starburst galaxies \citep[$D$=3.5 Mpc;][]{rekola2005}. NGC~253 is a nearly edge-on ($i=78.5$\degr) spiral with a strong bar that feeds its nuclear starburst \citep[e.g.,][]{scoville1985,peng1996,arnaboldi1995,engelbracht1998,das2001,paglione2004,meier2015,ando2017}. 
The central $\lesssim$200 pc region drives a galactic superwind to a distance of $\sim$10 kpc at hundreds of \kms  \citep{weaver2002,strickland2002,westmoquette2011,bolatto2013,walter2017}. The starburst region hosts two regions that could host the forming galactic center: the radio core with the brightest radio source in the galaxy \citep[``TH2''; e.g.,][]{turner1985,ulvestad1997}, and the IR core \citep[``IRC''; e.g.,][]{watson1996,kornei2009}. There are a number of candidate large star clusters 
detected as radio or sub-mm knots and associated molecular gas clumps, near the radio and IR core \citep{turner1983,turner1985,ulvestad1997,leroy2018,mangum2019}.  

The gas velocity fields of the nuclear region of NGC~253 are complex, resulting from combinations of distinct dynamical origin. 
The extended ionized gas takes the form of a wind that accelerates with distance from the galaxy \citep{westmoquette2011}. 
At radii $\sim$10\arcsec$\simeq$170 pc, the ionized gas kinematics are consistent with solid-body rotation of the nuclear disk, with a velocity gradient along the galaxy major axis PA$=$51\degr. However at the radio core and to a lesser extent the IR core, the velocity field is dominated by a distinct pattern with a gradient nearly perpendicular to the major axis \citep{anantha1996,das2001,rico2006}. 
These 
authors suggest that this central structure could trace outflow, an accreted object, or a secondary nuclear bar. 
The NGC~253 starburst powers a  molecular outflow
near the IR core 
\citep{weaver2002,strickland2002,bolatto2013,walter2017}.
Kinematic evidence for nuclear outflow near the IRC were seen in high resolution
Brackett $\gamma$ observations by \citet{gunthardt2015,gunthardt2019}; 
however, the location of the dynamical center of the galaxy remains ambiguous. 
 Severe dust extinction ($A_V\gtrsim 20$) greatly complicates kinematic analysis of optical-NIR gas tracers.
 
In this project, the 
goal was to determine 
the dynamics of ionized gas in the starburst nucleus of NGC~253 and  
the forming star clusters 
using an emission line that is bright and relatively impervious to extinction.  
To this end, we observed the nucleus of NGC 253 with NIRSPEC on Keck II, obtaining slit-spectra of the Brackett $\alpha$ emission at 10~\kms\ resolution.  Simultaneous imaging
of the slit with SCAM allow registration of 
the slits spatially with respect to near-IR continuum emission. 
 The resulting spectral cube covers the central $\sim$5 x 25\arcsec\ (90 x 360) region with a 
0\farcs5 slit, or 
$\sim 8.5~\rm pc$ 
at 3.5 Mpc.

\section{Observations and Data} \label{sec:data}

\begin{figure*}
\includegraphics[width=0.47\textwidth]{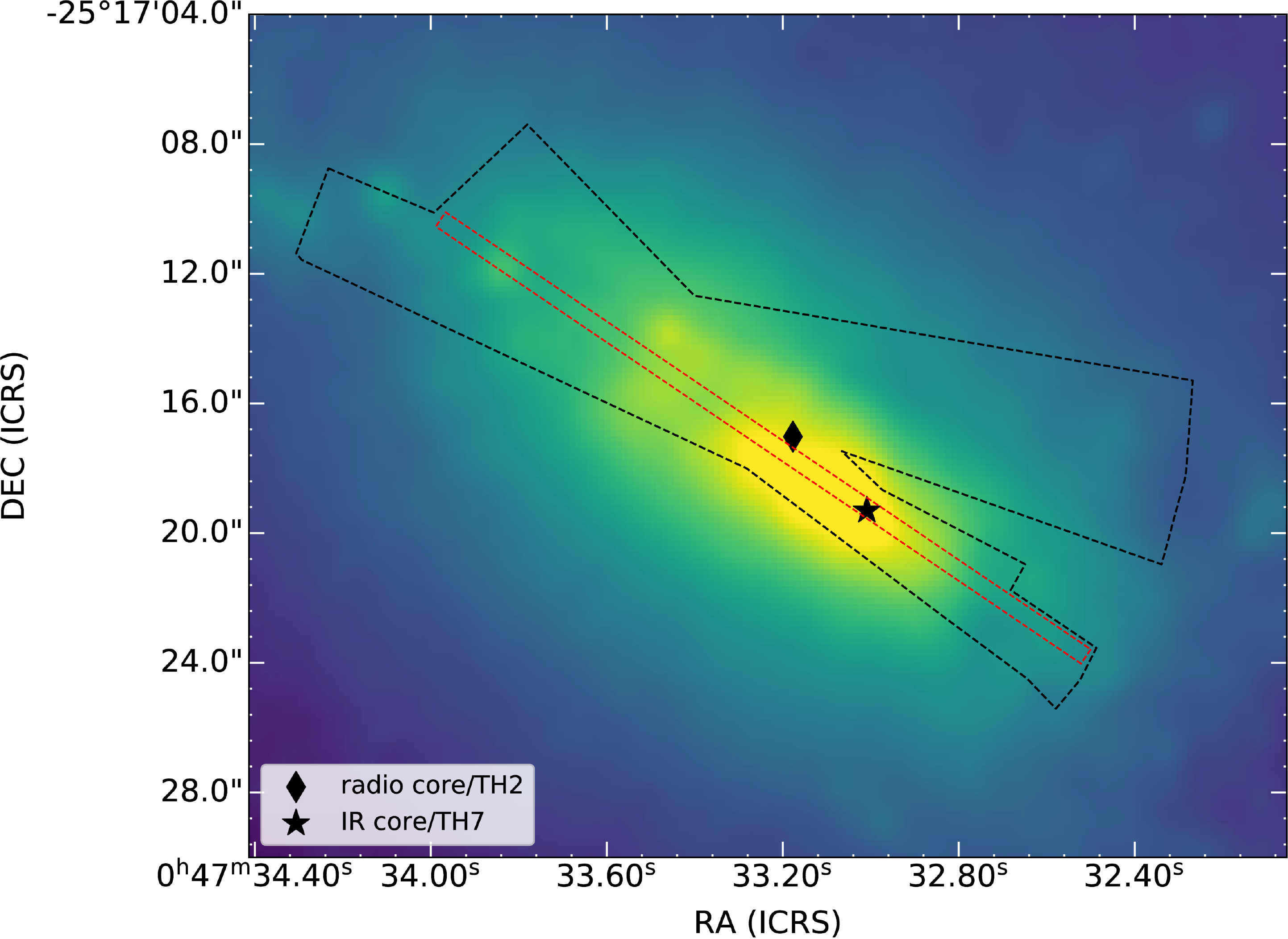}
\raisebox{8ex}{ % 
\includegraphics[width=0.48\textwidth]{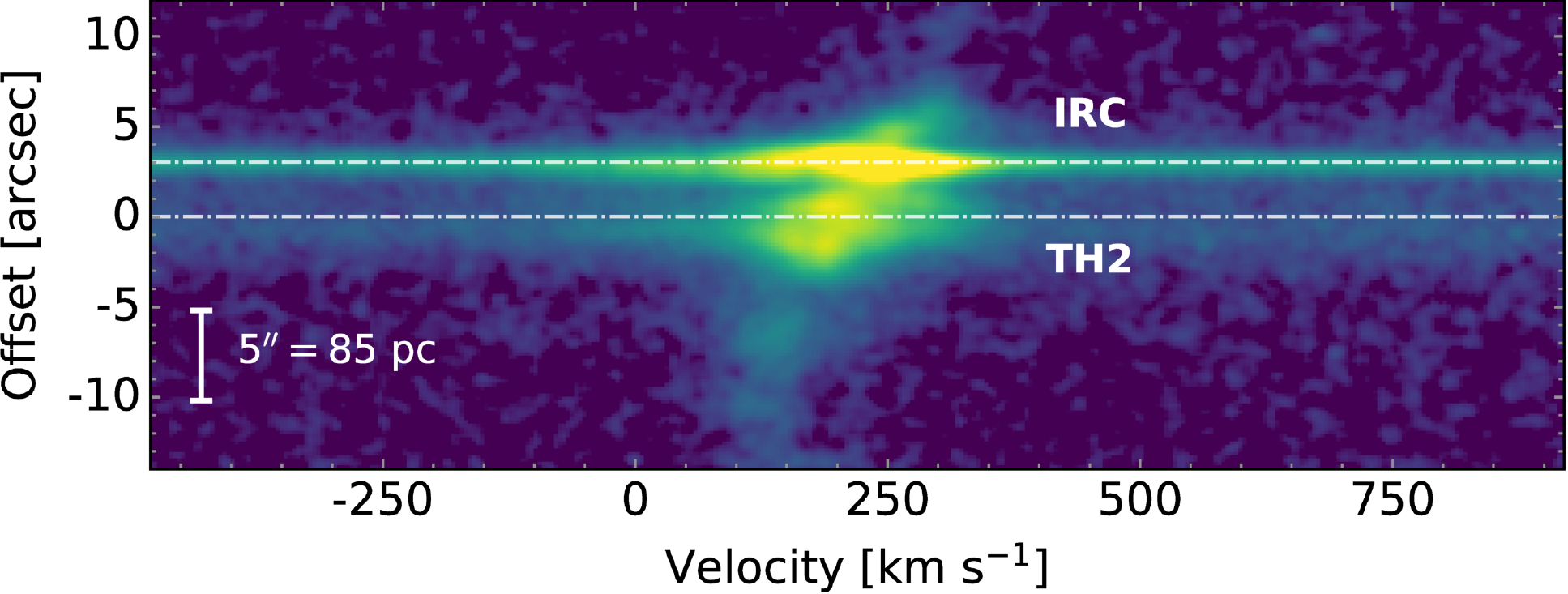}}
\centering
\caption{ NIRSPEC observations of the starburst in NGC~253. {\it Left}) Stacked SCAM image of the $K$-band continuum (color), with the region bounded by all slits in our map (black outline). The nonthermal source TH2 at the starburst's radio peak, along with the IR peak in the IRC, are marked by diamond and star symbols, respectively. {\it Right}) Reduced 2D echelle spectrum from the single, central slit position highlighted in red on the left. The spectrum runs from NE to SW in the positive vertical direction, {with the position of TH2 set as the origin. The brightest
\bra\ emission is detected at the radio core (near TH2) and at the IRC.} 
The spectrum displayed here was combined with the remaining 30 spectra acquired across the region to generate a \bra\ data cube (Sec.~\ref{sec:data}). \label{fig1} }
\end{figure*}

 We observed NGC~253 with NIRSPEC 
on Keck II \citep{mclean1998} in the first half-night on Dec 7, 2017. Observing parameters and 
properties of NGC~253 are reported in Table~\ref{tab:tab1}. 
Spectra were taken 
in high-resolution (echelle) mode in the KL band, using the 0\farcs432$\times$24\arcsec\ 
slit. The echelle and cross-disperser angles were set to 64.42{\degr} and 34.3{\degr}, respectively, yielding a wavelength coverage of 4.031-4.086 $\mu$m in the 19th echelle order. The seeing ranged between 0\farcs7 and 1\farcs1.

\begin{table}
 \centering
 \caption{Properties of NGC~253 and its two brightest nuclear sources, along with the NIRSPEC observing parameters.\label{tab:tab1}}
 \begin{threeparttable}
 \centering
 \begin{tabular}{ p{0.48\columnwidth} r} 
  \hline
  \hline
  Distance and angular scale & 3.5 Mpc; 1{\arcsec}=17 pc\tnote{(1)} \\
  Inclination of galactic disk $i$ & 78.5{\degr}\tnote{(2)} \\
  PA of disk major axis  & 51{\degr} \\
  PA of bar major axis / x$_1$ orbits  & 70{\degr} \\
  PA of bar minor axis / x$_2$ orbits & 45{\degr} \\ 
  Systemic velocity (heliocentric) $v_{\mathrm{sys}}$ & {$226$} {\kms}\tnote{(3)} \\
  Ionized galactic wind inclination & 12{\degr}\tnote{(4)} \\
  PA of galactic wind & 140{\degr} \\
  Opening angle of galactic wind & 60{\degr} \\
  Outflow speed of galactic wind  & $\gtrsim$100-300 {\kms} \\
  Predicted SMBH mass $M_{\mathrm{BH}}$ & $\sim 2\times10^7$ {\msun}\tnote{(5)} \\
  \hline
  Radio center, source TH2 (ICRS) & $00^{\mathrm h}47^\mathrm{m}33\fs18$, $-25{\degr}17\arcmin16\farcs94$\tnote{(6)} \\
  Radio core mass $M_{\mathrm{dyn}}^{\mathrm{TH2}}$ & $\sim7\times10^6$ {\msun}\tnote{(7)} \\
  Infrared center, IRC (ICRS) & $00^{\mathrm h}47^\mathrm{m}32\fs99$, $-25{\degr}17\arcmin19\farcs74$\tnote{(8)} \\
  IRC mass $M_{\mathrm{dyn}}^{\mathrm{IRC}}$ & $\sim 5\times10^5$ {\msun}\tnote{(9)} \\
  \hline
  Observing wavelength & $\lambda_{\mbra} = 4.052$ $\mu$m \\
  Slit size &  0\farcs432$\times$24{\arcsec} \\
  Echelle angle & 64.42{\degr} \\
  Cross-disperser angle & 34.3{\degr} \\
  Seeing & 0\farcs7-1\farcs1 \\
  Velocity resolution & 12 {\kms} \\ 
  Mapping Area &  $\sim$18{\arcsec}$\times$30{\arcsec}\tnote{(10)}  \\
  Total exposure time  &  64 min\tnote{(11)} \\
  Peak pixel S/N ratio & 7-72\tnote{(12)} \\
  \hline
  \hline
 \end{tabular}
 \begin{tablenotes} \footnotesize
  \item[(1)] \citet{rekola2005}.
  \item[(2)] Galactic disk and bar properties adopted from \citet{das2001}, \citet{paglione2004} and references therein. 
  \item[(3)] {This paper, Sec.~\ref{subsec:kin1}}.
  \item[(4)] Ionized wind properties from \citet{westmoquette2011} modeling of H$\alpha$ outflow cone. 
  \item[(5)] Predicted using $M$-$\sigma$ relation \citep{combes2019} with measured NGC~253 bulge stellar velocity dispersion from \citet{oliva1995}. The presence of a SMBH in NGC~253 is expected but unconfirmed.
  \item[(5)] Coordinates of radio peak in NGC 253, identified as nonthermal source TH2 in \citet{turner1985}. 
  \item[(6)] Estimate from \citet{rico2006} within $r<7$ pc of TH2, using H92$\alpha$ mapping with a $\lesssim$0\farcs5 beam. 
  \item[(7)] Coordinates of IR peak (IRC) from \citet{leroy2018}, their source \#5. Coincident with thermal source TH7 \citep{turner1985}. 
  \item[(8)] Virial mass estimate from \citet{leroy2018} using the resolved source size of 2.1 pc.
  \item[(9)] Size of the NIRSPEC mapping region, with 31 slit positions. 
  \item[(10)] Combined NIRSPEC exposure time for all spectra across the nuclear starburst. 
  \item[(11)] Signal-to-noise ratio (S/N) of the brightest pixel in echelle spectra.
 \end{tablenotes}
 \end{threeparttable}
\end{table}

We obtained a total of 31 120s exposures at slit positions spanning the central $\sim$$18\arcsec\times30\arcsec$ of the nuclear region, shown in Figure~\ref{fig1}. We initially orientated the slit along the galaxy's major axis, at PA$\simeq$45{\degr}. The slit PA drifted with Earth's rotation such that the slits for our final exposures were oriented at PA$\simeq$80{\degr}. Sky spectra were acquired by offsetting away from NGC~253 after the 8th science exposure and again after the final science exposure. Two calibration stars were observed prior to and after completing  all NGC~253 observations (HD225200 and HD12365, respectively). Images of the slit on the sky in the K band were simultaneously acquired using the NIRSPEC Slit-Viewing Camera (SCAM). These images allowed for  
registration of the slits with respect to the near-IR background. In this way we were able to build a cube.

Each slit position yielded a 2D echelle spectrum which we rectified, reduced, and calibrated, using our calibration star spectra along with arc lamp spectra acquired at the beginning of the night. We first subtracted raw NIRSPEC images by sky spectra, divided by a median-normalized flat-field image, and iteraitvely removed hot/cold pixels. We then spatially and spectrally rectified the images using an \texttt{Python}-based implementation the  
\texttt{REDSPEC} reduction code\footnote{\url{https://www2.keck.hawaii.edu/inst/nirspec/redspec.html}}. The reduced echelle spectra have spatial information in the vertical axis (along the slit) and spectral information in the horizontal axis, as shown by the example spectrum in Fig~\ref{fig1}. Each pixel has a size of 5.482$\times 10^{-5}$ $\mu$m along the spectral axis and 0\farcs192, the nominal value, along the spatial axis. The spectral resolution is about 3 pix $\simeq$ 12 {\kms}. 

\subsection{The NIRSPEC Cube} \label{subsec:cube}

To take full advantage of our data set, we combined the separate spectra into a spectral cube, with RA/DEC on the $x$- and $y$-axes and wavelength/velocity along the $z$-direction. Constructing the data cube involved mapping between pixels in each 2D spectrum and pixels on the cube grid, requiring astrometric {registration}  of the slits. Registration 
was performed by creating a stacked, slit-free SCAM image in which a handful of IR sources are well-detected (Fig.~\ref{fig1}). To create the stacked image, we sky-subtracted, rotated, and aligned all individual SCAM exposures (showing the slit trace) to a common reference image. We then masked out the slit trace in each image and median-combined all images of acceptable quality. During this process we tracked the position of each slit, shifting and rotating to the reference image. Once the stacked image was obtained, we measured the pixel positions 
of detected K-band sources, and matched them with sources with known sky coordinates detected in an archival HST F160W image (Proposal ID 12206, PI Westmoquette), yielding registration of
the SCAM image and HST images accurate to $\lesssim$0\farcs1 (rms error). 

Inferring the pixel mapping for each spectrum requires identification and measurement of a reference point, in addition to the sky positions of the slits (the center of the slit does not correspond to the center of the rectified spectrum). As such, we measured the pixel positions of \bra\ sources in each spectrum that we could match to known sources detected in the NIR imaging, and established the reference pixel required to map spectrum pixel coordinates to sky coordinates.

After registration, 
we constructed the NIRSPEC cube in the following way. First, we interpolated each echelle spectrum across the slit width, assuming a constant light profile in that direction, and then mapped the data from each pixel in each wavelength slice onto the coordinate frame of the cube. We median-binned the mapped data to a regular grid, and applied smoothing with a Gaussian kernel to remove small-scale artifacts resulting from regions with sparse data. Finally, we interpolated each image slice to the final grid with a spatial scale of 0\farcs12 pix$^{-1}$. The spatial resolution of the cube is $\simeq$1\farcs0, approximately the same as in individual spectra.  A map of the Brackett line emission was generated by fitting continuum-only pixels 
with a first order polynomial and subtracting this contribution from the data cube. The \bra\ cube is shown with channel maps in Figure~\ref{fig2}.

\section{RESULTS} \label{sec:results}

\begin{figure*}
\includegraphics[width=0.75\textwidth]{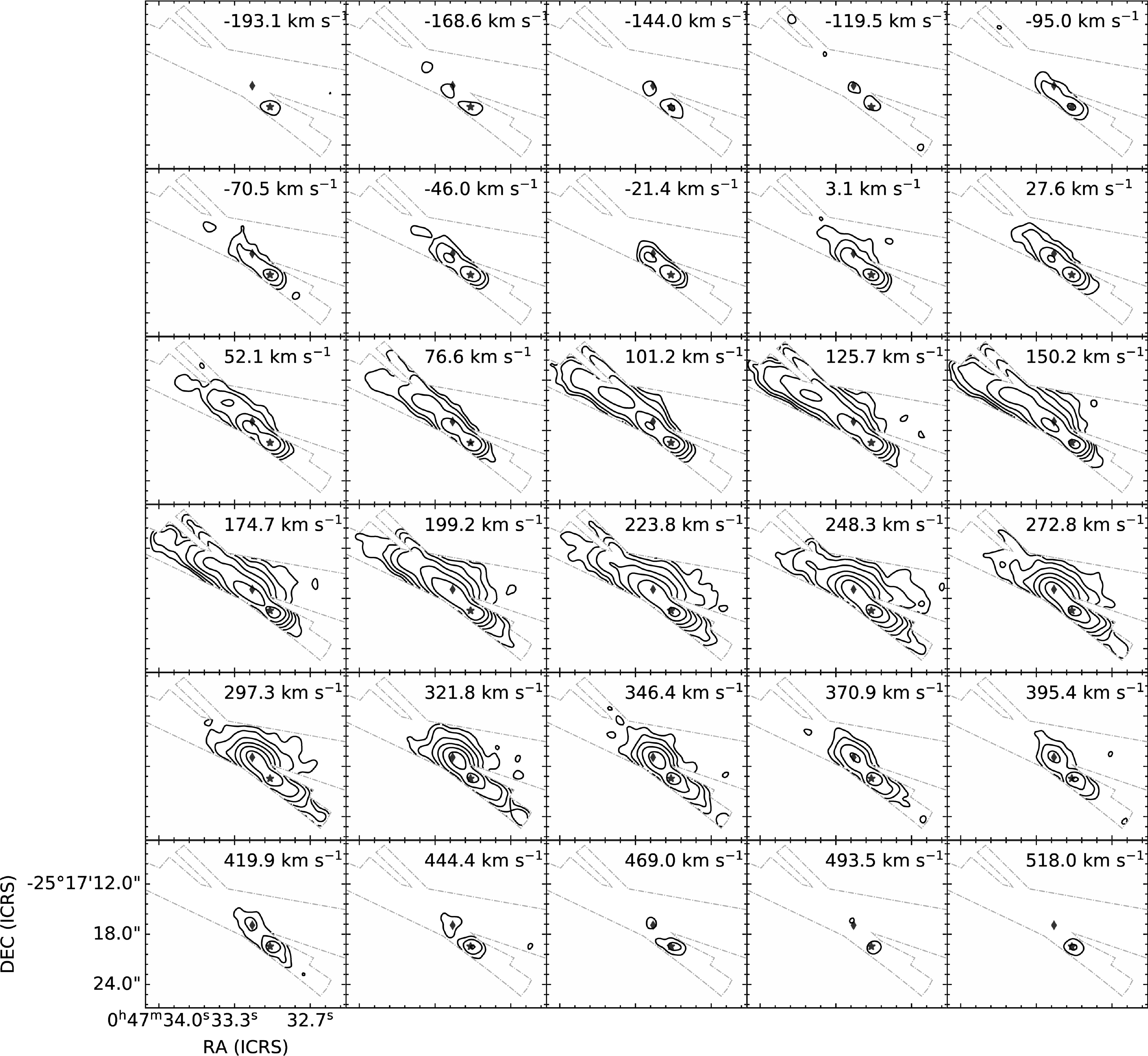}
\centering
\caption{Channel maps of the NIRSPEC Br $\alpha$ cube. To produce this figure, channels are bin-averaged by 6 pixels, to a width of $\Delta v=24.5$ {\kms}. Contours show $\sigma_{\mathrm{rms}}\times 2^{n/4}$, $n={6,10,14,18,...,44,48}$, where $\sigma_{\rm{rms}}$ is the rms noise in channels away from the \bra\ line. As in Fig.~\ref{fig1}, the IRC and TH2 are marked as the star and diamond symbol, respectively. Four \bra\ sources are identifiable in these maps: a bright peak near IRC that has a broad component visible across all channels, another bright/broad source at the radio core near TH2 with an apparent gradient, a third clump of narrower emission to the NE of TH2 (apparent at 101.2 {\kms}), and a fourth clump towards the NE corner of the FoV (apparent at 52.1 {\kms}). 
\label{fig2}}
\end{figure*}

\begin{figure}
\centering
\includegraphics[width=0.95\columnwidth]{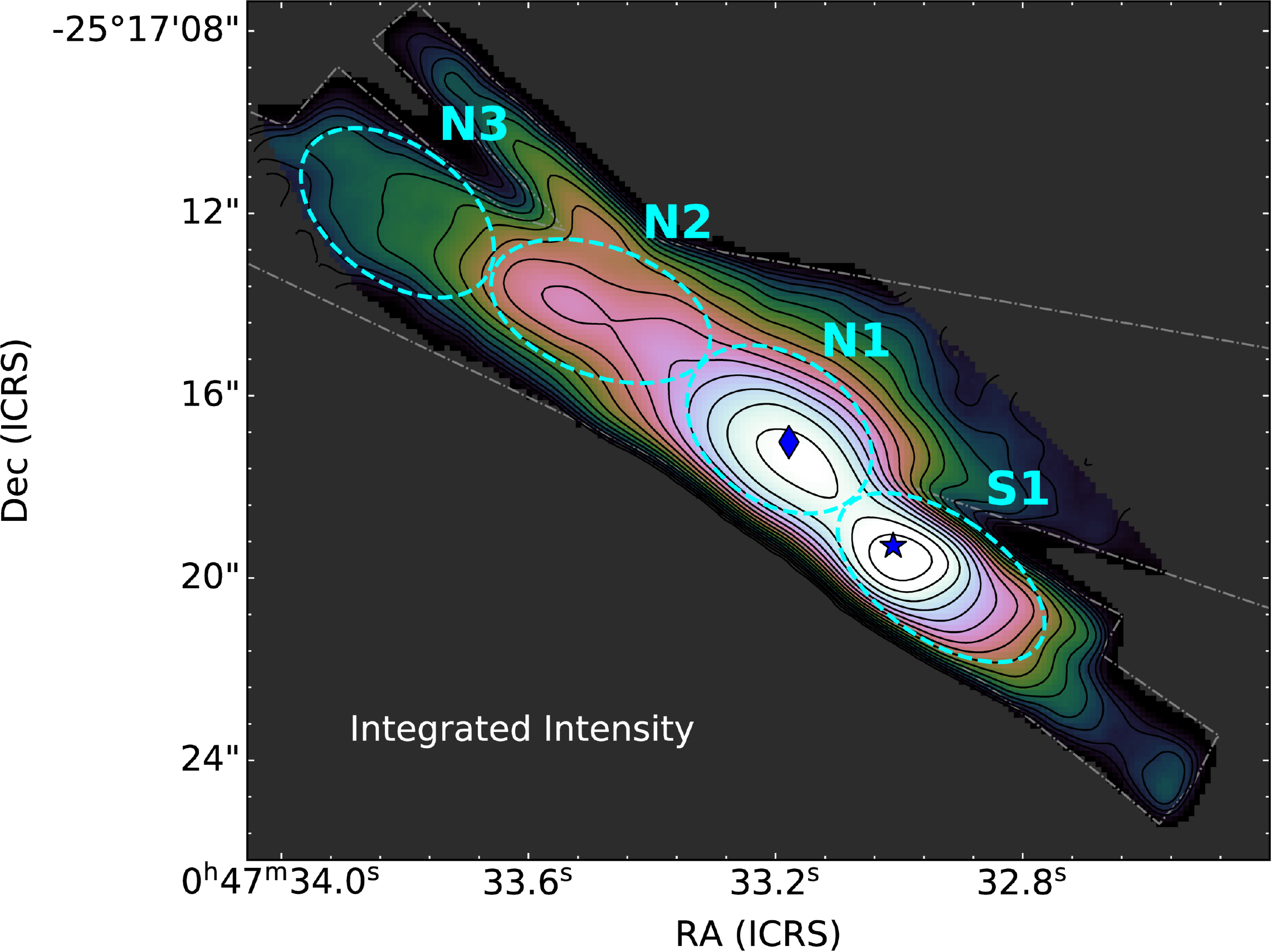}
\caption{Velocity-integrated Brackett $\alpha$ emission (line flux map) in the core of NGC~253, generated from the NIRSPEC cube by summing pixels at $>$1.5$\sigma_{\mathrm{rms}}$ along the spectral axis. The slit coverage is indicated by the grey outline, and the IR core and TH2 are again marked as the star and diamond symbols, respectively. 
We detect four likely distinct Brackett sources: N3, N2, N1, and S1 (dashed cyan ellipses). The elliptical apertures shown here are used to extract the total spectrum of each source.
\label{fig3}}
\end{figure}

The cube of Br $\alpha$ emission in the starburst core in NGC 253, shown in Figure \ref{fig2}, reveals  
four sources within our slit coverage. We identify these sources as S1, N1, N2, and N3 from the SW to NE along the major axis, and label them on the velocity-integrated Br $\alpha$ intensity map in Figure~\ref{fig3}.

Source S1 is {at the IR core (IRC), a region hosting the starburst's brightest NIR source, a $\sim$6 Myr old, $M\sim 10^6$ {\msun} SSC \citep[e.g.,][]{watson1996,kornei2009,ontiveros2009,davidge2016,gunthardt2019}. }
Several other candidate SSCs are identified near the bright SSC as fainter thermal IR-radio continuum knots \citep[e.g.,][]{ontiveros2009,gunthardt2015,leroy2018}. Broad recombination lines suggest an ionized outflow driven by the SSC formation near the IRC
\citep[][]{gunthardt2019}. {We assume Brackett source S1 primarily traces the HII region excited by the bright SSC in the IRC, with small contributions from fainter HII regions around nearby clusters.} 

 Source N1 is associated with the radio core, which contains the galaxy's brightest radio source, TH2 \citep{turner1985}. TH2  {is characterized by non-thermal radio continuum with brightness temperature $\gtrsim$50,000 K \citep{turner1983,turner1985,ulvestad1997}, and no clear infrared counterpart \citep{gunthardt2015}, thus is unlikely to be
 an HII region}. While its origin remains unclear, the bright radio source TH2 is potentially linked to the presence of a supermassive black hole \citep{turner1985,muller2010}. Brackett source N1 should also trace candidate SSCs and their HII regions within the radio core, identified as compact thermal IR-radio sources \citep[e.g.,][]{turner1985,ulvestad1997,leroy2018,mangum2019}. {The line profile of ionized gas within the radio core, only possible with spectroscopy at $K$ band or longer wavelengths, has been measured for RRLs \citep{rico2006,kepley2011,bendo2015} and IR emission lines \citep[e.g.,][]{rosenberg2013,gunthardt2015}.}
 
N2 and N3 comprise weaker \bra\ emission compared to N1 and S1, and are possibly linked to \citet{gunthardt2015} sources A5-6 and A3/10, respectively. Optical source `spot {\it a}'  from \citet{watson1996} might also contribute to the emission near N2 and N3. 

\subsection{Brackett Line Profile} \label{subsec:lineprofile}

\begin{figure*}
\begin{center}
\includegraphics[width=0.7\textwidth]{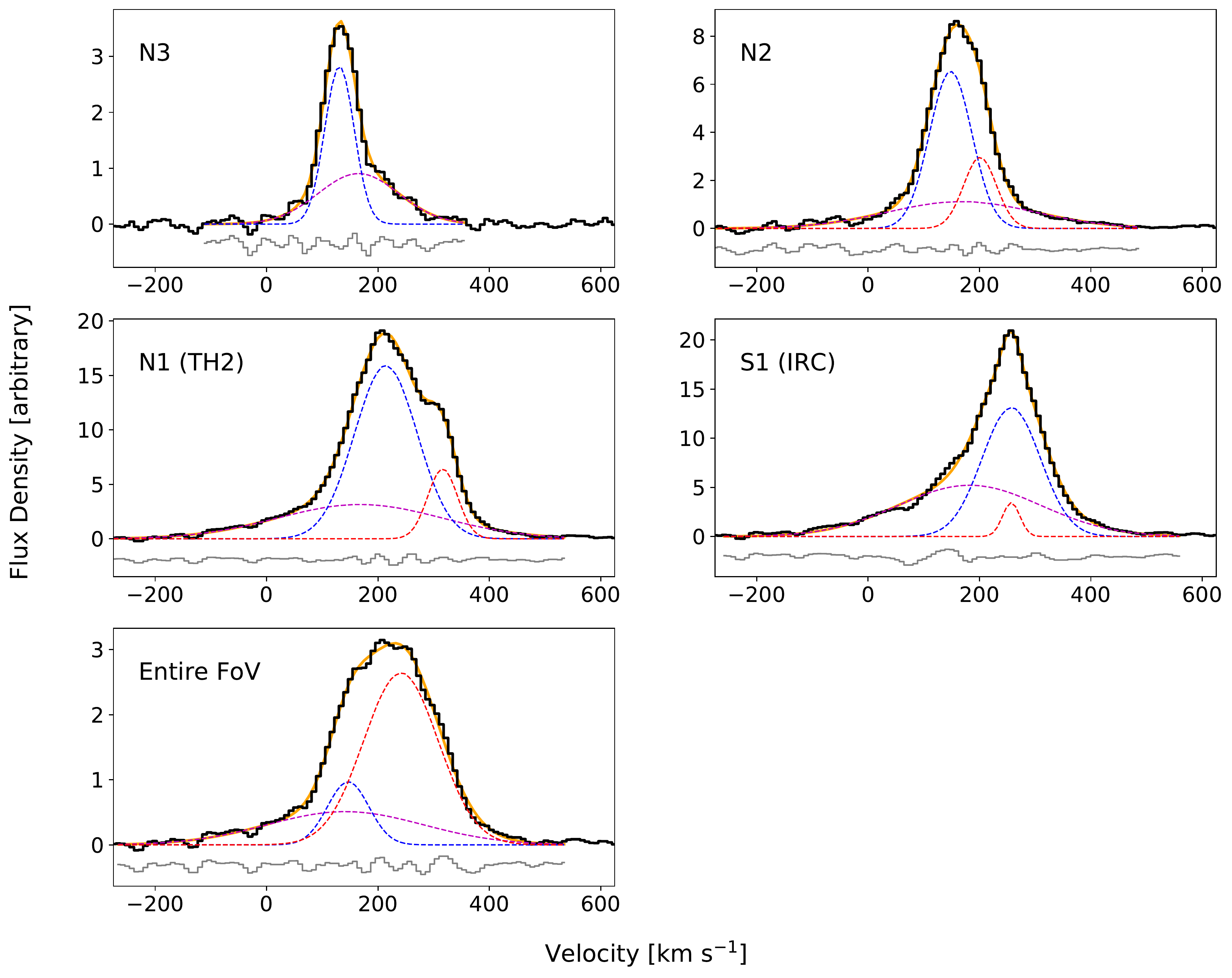}
\caption{\bra\ spectrum of identified sources {(along with the entire imaged region)} and best-fit Gaussian models of the line profiles. Spectra (black line) were extracted by summing all pixels within apertures (Fig.~\ref{fig3}) along the spatial dimensions of the \bra\ cube. Intensities in all panels are scaled by a common normalization factor. With clear asymmetries and structure in the line profiles, none of the sources can be modeled well with a single-component Gaussian profile. Models with two or more Gaussian components provide much better fits to the data, as shown in each panel by the best-fit model (solid orange line), its individual fit components (dashed lines), and the fit residuals (grey curve). All sources exhibit a ``primary" bright peak component and a ``broad" component that may appear as a blue wing extending out to large negative velocity away from line center with weak, if any, corresponding emission on the red side. A third, ``narrow" and red component is additionally required to model N2, N1, and S1. \label{fig4}}
\end{center}
\end{figure*}

Figure~\ref{fig4} shows the 
Br $\alpha$ profiles for S1, N1, and N2, {N3} averaged over the $\sim$$2\farcs4\times 1\farcs4$ apertures in Fig.~\ref{fig3}. To characterize the line profile, we first fit each spectrum with an oversimplified model of a single-component Gaussian profile and infer velocity centroids and line widths. We also estimated {\bra} equivalent width, a proxy for age in young massive clusters, by dividing these models by the best-fit models of the continuum (Sec.~\ref{subsec:cube}) and integrating over the line. The single-component models reveal a shift in \bra\ velocity centroid by roughly $+$120 {\kms} from N3 to S1
along with broad linewidths reaching FWHM$\sim$200 {\kms} near N1 and S1. Equivalent width is largest for N1, $W(\mbra)\sim 250${\AA}, and decreases to $W(\mbra)\sim 100-150${\AA} for S1, N2, and N3. 

More realistic models of the {\bra} line comprising two or three Gaussian components rather than one, selected using the Bayesian Information Criterion, are shown in Fig.~\ref{fig4}. Parameters for each component of these models are reported in Table~\ref{tab:tab2}.
Generally the profile is characterized by a broad component (FWHM$_{\mathrm{broad}}\sim300$-350 {\kms}) at the base of a primary emission peak (FWHM$_{\mathrm{primary}}\sim90$-130 {\kms}). The broad emission has the highest intensity and largest linewidth for sources {N1 and S1 at the radio and IR cores}, respectively, where it appears almost exclusively on the blue side of the line. A blue wing is similarly exhibited by the Br $\gamma$ line in this region \citep{gunthardt2019} and by H$\alpha$ tracing more extended gas in regions of low extinction \citep{westmoquette2011}. Blue wings are common in nebular lines from starburst regions, and suggest that
extinction may be blocking the red wing from view, so these linewidths are lower limits.
In addition to the broad and primary components, the line shows a third, narrow (FWHM$_{\mathrm{narrow}}\sim60$ \kms) component strongest near N1 with a peak reaching an offset of $\Delta v_{\mathrm{n,p}} \equiv v_{\mathrm{narrow}} - v_{\mathrm{primary}} \simeq +100$ \kms\ from the primary component. The Br $\gamma$ line shows evidence of this
narrow feature, however it is not identified as a distinct peak \citep{gunthardt2019}.

\begin{table*}
\begin{center}
\caption{Properties of \bra\ sources identified in Fig.~\ref{fig3}. Sky coordinates are based off of best-fit 2D profiles, while kinematic properties are from the best-fit Gaussian models shown in Fig.~\ref{fig4}. \label{tab:tab2}}
\begin{threeparttable}
\begin{tabular}{lccccc} 
		\hline
		\hline
		Source & 0$^{\mathrm h}$47$^\mathrm{m}$~(ICRS)\tnote{(1)} & -25{\degr}17\arcmin~(ICRS)  & $v_{\rm primary}$ [\kms]\tnote{(2)} &  FWHM$_{\rm primary}$ [\kms]\tnote{(3)} & $F_{\rm primary}/F_{\rm tot}$\tnote{(4)}\\
		\hline
		N3 & 33\fs83 & 12{\arcsec} & $131\pm2$ & $62\pm7$ & $0.53$ \\
		N2 & 33\fs52 & 14{\arcsec}  & $148\pm14$ & $89\pm14$ & $0.51$ \\
		N1 (radio core) &  33\fs16 & 17{\arcsec}  & $214\pm1$ & $133\pm4$ & $0.59$ \\
		S1 (IRC) &  33\fs01 & 19{\arcsec}  & $257\pm1$ & $123\pm4$ & $0.49$ \\
		\hline
		\hline
		\hline
		$v_{\rm broad}$ [\kms]\tnote{(5)} & FWHM$_{\rm broad}$ [\kms]\tnote{(6)}   & $F_{\rm broad}/F_{\rm tot}$\tnote{(7)}  & $v_{\rm narrow}$ [\kms]\tnote{(8)}   & FWHM$_{\rm narrow}$ [\kms]\tnote{(9)}   & $F_{\rm narrow}/F_{\rm tot}$\tnote{(10)}\\
		\hline
		
		$165\pm13$ & $170\pm23$ & $0.47$  & ... & ... & ... \\
		$168\pm11$ & $313\pm35$ & $0.30$  & $201\pm14$ & $69\pm14$ & $0.18$ \\
		 $168\pm6$ & $347\pm14$ & $0.30$  & $316\pm1$ & $63\pm2$ & $0.11$ \\
		$182\pm5$ & $301\pm7$ & $0.47$  & $257\pm1$ & $36\pm4$ & $0.04$ \\
		\hline
\end{tabular}
\begin{tablenotes}
\item[(1)] Source centroid coordinates have a 1-$\sigma$ uncertainty of $\sim$1\arcsec.
\item[(2),(5),(8)] Centroid velocity {(heliocentric)}  for each Gaussian component of the best-fit two- or three-component model. The components are labeled as ``primary'', ``broad'', and ``narrow" (the narrow component can be fit to all but source N3). 
\item[(3),(6),(9)] FWHM of each Gaussian component in the best-fit model.
\item[(4),(7),(10)] Fraction of total line flux in each component of the best-fit model (ratio of each component flux that to the summed flux of all components).
\end{tablenotes}
\end{threeparttable}
\end{center}
\end{table*}

\begin{figure*}
 \begin{center}
\includegraphics[width=0.8\textwidth]{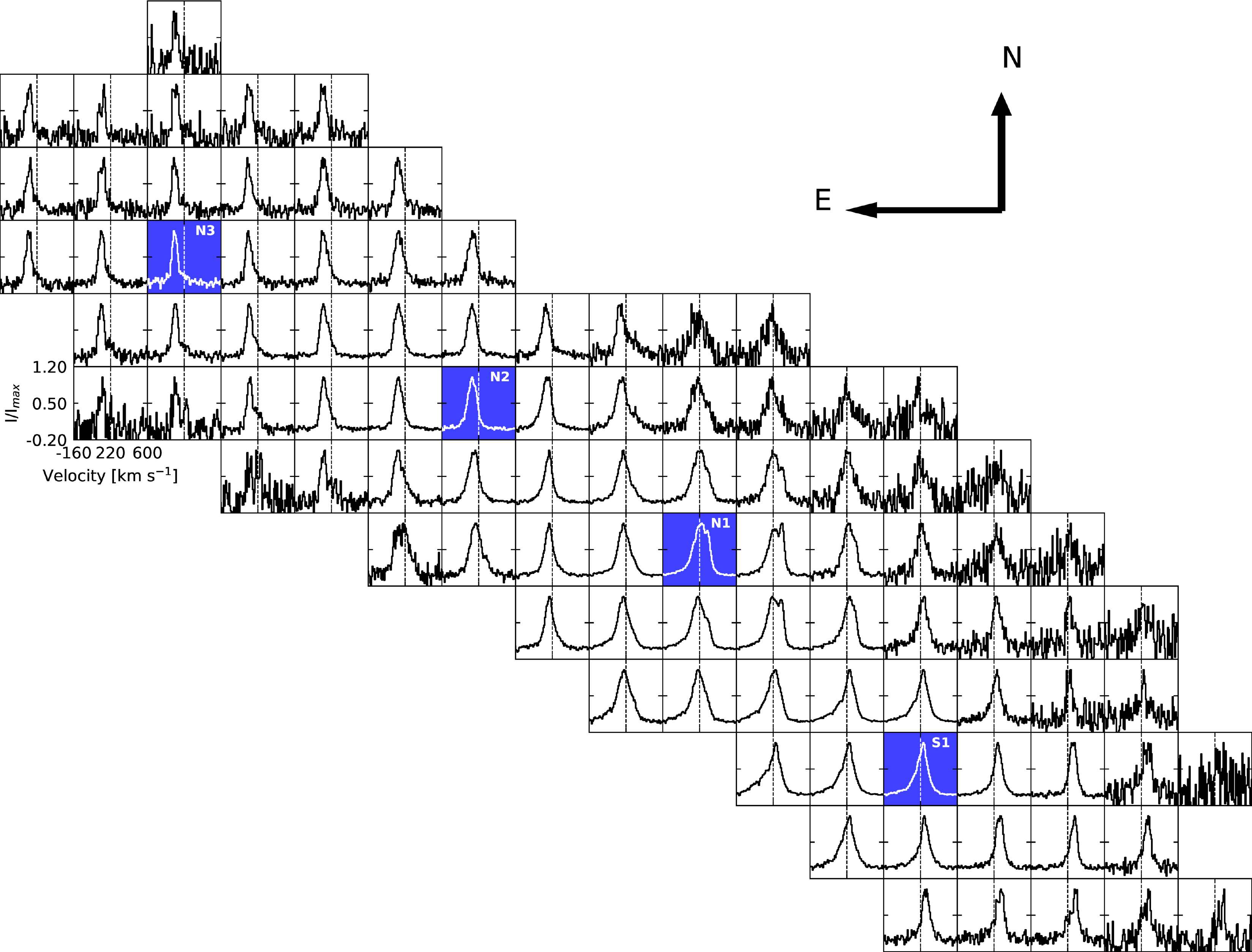}
\caption{Map of \bra\ line profile across the NIRSPEC FoV. Each panel gives the summed spectrum within a $1\farcs2\times 1\farcs2$ box centered at that location, and each of these spectra are individually normalized so that variation of the line shape across the region is clear (line peak intensities should not be compared between panels). Blue-highlighted panels show the location of \bra\ sources. The velocity axis is identical for all spectra, with a vertical line marking the heliocentric systemic velocity $v_{\mathrm{sys}}=220$ {\kms}. Kinematic patterns are clear, in particular: an increase in strength of the broad blue component towards the SE, a double-peaked line structure just east of TH2, and a bulk shift in the line peak velocity centroid to higher velocity (redshift) from NE to SW, along the major axis. 
\label{fig5}}
\end{center}
\end{figure*}

{While insightful, the average line profiles (Fig.~\ref{fig4}) blend together systematic motions within the $\sim$2\arcsec apertures used for extraction}. Figure~\ref{fig5} shows the line profile mapped across the central region. 
This map reveals a major axis velocity shift consistent with the change in line centroid between the sources, with velocity increasing (redshifted) from NE to SW. The picture is most complicated near N1, where the narrow third component is strongest. Interestingly, {$\sim0\farcs8$ east of TH2, the narrow red component reaches a peak intensity that is $\sim$1.4$\times$ higher than that of the the primary component at that location}. This narrow red feature could be a kinematically distinct source or a portion of a larger gas flow. 
We note that the irregular line profile near TH2 suggests the possibility of $>$3 components that would require higher resolution to distinguish and characterize.

The spectrum map (Fig.~\ref{fig5}) also show that the broad blue wing increases in strength to the south and east of the peak positions in both N1 and S1, with a peak just east of S1, where the feature contributes $\sim$50\% of the line flux. 
The velocity centroid of broad emission reaches a minimum of $\Delta v_{\mathrm{b,p}} \equiv v_{\mathrm{broad}}-v_{\mathrm{primary}}\simeq -90$ \kms\ {at a distance of 1-$1.5\arcsec$ around the eastern side of the IRC, corresponding to the panels immediately to the left of S1 in Fig.~\ref{fig5}. Moving NE away from S1, the offset shifts positively, reaching $\Delta v_{\mathrm{b,p}}\sim 0$ \kms\ in the region north of N1 and west of N2 (in Fig.~\ref{fig5}, panels above N1 and to the right of N2). Further NE, the broad component becomes a {\it red} wing with $\Delta v_{\mathrm{b,p}}\gtrsim 0$ \kms\ near N2 and N3 (any panels to the left of N2 in Fig.~\ref{fig5}).} This change in $v_{\mathrm{b,p}}$ is due to the major-axis shift in the primary component, while the broad component centroid remains constant to within $\lesssim$20 {\kms}. {Nonetheless, the broad component shows a clear blueshift to the SE of N1 and S1, explaining the $-90$ {\kms} offset near the IRC.
 Heavy extinction near the IRC and radio core can explain the much of the blue-red asymmetry of the broad feature; the observed blue wing may not reflect the full kinematic structure.}

\subsection{Nuclear Velocity Structure} \label{subsec:kin}

\begin{figure}
\begin{center}
\includegraphics[width=0.95\columnwidth]{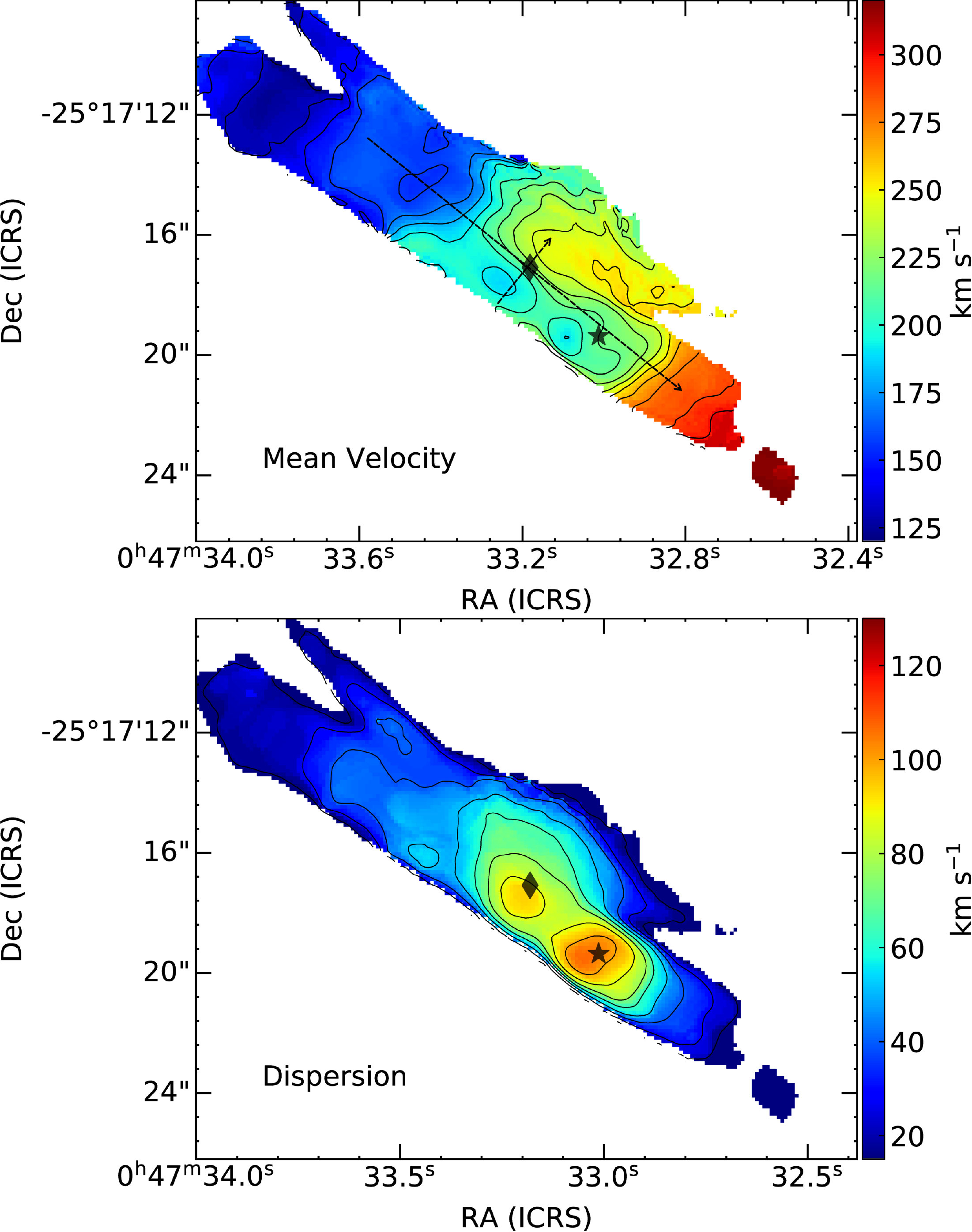}
\caption{Velocity field of the nuclear region in NGC~253. The upper and lower panels show the intensity-weighted mean velocity (first moment) and dispersion (second moment, equivalent to the linewidth $\sigma_v$), respectively. Both maps were generated from the NIRSPEC cube using pixels detected at $>6\sigma$. {Contours mark intervals of 12 {\kms} and 10 {\kms} in the upper and lower panels, respectively.} As in other figures, the star marks the IRC and the diamond marks the radio peak TH2. The dashed arrows mark pseudoslits used to generate major- and minor-axis PV images.
\label{fig6}}
\end{center}
\end{figure}

The NIRSPEC cube (Fig.~\ref{fig2}) allows us to probe the detailed ionized gas kinematics using the standard methods of spectral cube analysis. 
First and second moment maps, representing the intensity-weighted line-of-sight velocity field and dispersion, respectively, were generated from the \bra\ cube using pixels at >6$\sigma_{\mathrm{rms}}$ to identify distinct patterns in the velocity structure (Figure~\ref{fig6}).

\subsubsection{Major-axis Velocity Gradient} \label{subsec:kin1}

The velocity field in Fig.~\ref{fig6} is complex, comprising several distinct kinematic structures rather than a single ordered pattern. 
The simplest of these structures is 
apparent 
as the velocity contours at the NE and SW edges of the field (away from TH2 and IRC). Perpendicular to the major axis, these contours  indicate a positive gradient from NE to SW across the central $\sim$12{\arcsec}$\simeq$200 pc (projected). This signal is washed out near N1 and S1, where other kinematic structures clearly dominate the velocity field.

To further probe the rotation curve, we generated a PV image using a pseudoslit centered on TH2 oriented parallel to the major axis, at PA=51{\degr}. The resulting major-axis PV diagram (Fig.~\ref{fig7}, left) more clearly exhibits solid-body rotational pattern across the regions of more complex kinematic structure.  
The major-axis rotation curve, or 1D velocity profile, is derived from best-fit models of multi-component Gaussian profiles to the spectrum in each column of the PV image. We take the primary fit component (as opposed to the broad wing or red narrow component) to generate the curve shown 
 in the right-hand panel of Fig.~\ref{fig7}.   
The best-fit velocity gradient is 
10 {\kms} arcsec$^{-1}$.  
A major-axis gradient of this magnitude has been well-measured with many gas tracers, including
in  RRLs \citep{anantha1996,rico2006}, Br $\gamma$, H$_2$, and CO \citep{engelbracht1998,gunthardt2015,leroy2015}. 

From Fig.~\ref{fig7} it is clear that the major-axis velocity profile of the primary \bra\ component is more complex than a simple gradient from solid-body rotation. The residuals show a sinusoidal structure seemingly associated with sources; the largest residual peaks slightly precede positions of N1/TH2 and S1/IRC. Best-fit sinusoidal curves indicate residual amplitudes of 17 {\kms} and 12 {\kms} for N1 and S1, respectively.  
For both sources, residual peaks are offset by about $0\farcs5$ NE of the emission peaks (TH2 and IRC). This results in velocity gradients across the radio and IR cores of approximately $-20$ {\kms} arcsec$^{-1}$ and $-15$ {\kms} arcsec$^{-1}$, flipped with respect to the underlying 10 {\kms} arcsec$^{-1}$ gradient. Similar major-axis velocity residuals were reported in \citet{anantha1996} and \citet{gunthardt2015}. 

 The kinematic center, systemic velocity, and position angle of the nuclear disk rotation are estimated using BBarolo \citep{tedoro2015} to fit a tilted-ring model to the velocity field, with inclination fixed to $i=78.5^\circ$. The best-fit model has center ($\alpha_{\mathrm{icrs}}$, $\delta_{\mathrm{icrs}}$)=(00$^\mathrm{h}$47$^\mathrm{m}$33\fs09, $-$25$^\circ$17\arcmin17\farcs8) with an uncertainty of $\sigma \simeq 1\arcsec$, $v_{\mathrm{sys}}=226\pm11$ {\kms}, and PA $\phi_d=51\pm6${\degr}. {The disk PA and systemic velocity are in close agreement with the literature \citep[e.g.,][]{anantha1996,das2001}}. The dynamical center is within $\lesssim 0\farcs7$ of previous determinations \citep{anantha1996,muller2010,rosenberg2013}, and is $\simeq$1\farcs5 closer to TH2 than to the IRC. We note that tilted-ring modeling is limited in its ability to explain the nuclear kinematics, due to the significant non-rotational structure present in the velocity field.

\subsubsection{Substructure of the Radio Core and Source N1} \label{subsec:kin2}

The most prominent pattern in the velocity field (Fig.~\ref{fig6}) comprises the S-shaped velocity contours {at source N1}, centered close to the radio peak TH2. This kinematic substructure has a characteristic size of $\sim$2-3{\arcsec}$\simeq$30-50 pc, based on the contours. Unlike the underlying solid-body rotation pattern, the contours of N1 are nearly parallel to the major-axis, implying a velocity gradient of $\sim$+50 {\kms} arcsec$^{-1}$ from SE to NW along the minor axis. 
 Again, the \bra\ velocity structure is consistent with prior measurements of RRLs:  \citet{anantha1996} find a minor-axis gradient of $\sim$20-30 {\kms} arcsec$^{-1}$ for H92$\alpha$, \citet{rico2006} similarly find a 25 {\kms} arcsec$^{-1}$ for H92$\alpha$ but infer a larger value of 42 {\kms} arcsec$^{-1}$ for H53$\alpha$, closer to our estimate. {The $K$ band velocity fields presented in \citet{rosenberg2013} are the closest comparison to our \bra\ field; their H$_2$ field resolves the same three substructures identified in our map. Interestingly, their stellar velocity field also shows the S-shaped velocity contours, suggesting a common dynamical origin for the gas and stars in the radio core.}

 In the major-axis PV image (Fig.~\ref{fig7}, left), source N1 is {resolved into two bright components connected by fainter emission surrounding a central ``hole".} 
The bright components here were identified as the primary and narrow red component in the \bra\ line profiles (Fig.~\ref{fig4}).  {The pv ``hole" is centered at a position $\simeq$0.7-0.8\arcsec\ to the SW of TH2 and heliocentric velocity $v\simeq 230$ \kms, with an extent of $\Delta x\sim 2\arcsec$ along the position axis and 
 $\Delta v \sim$100 {\kms} along the velocity axis}. This structure is likely related to the local maximum in velocity dispersion near TH2 (Fig.~\ref{fig6}), which reaches $\sigma_v\simeq 90$ \kms. The major-axis PV morphology suggests that the radio core can be characterized by a single dynamical structure.
 
A PV image was extracted along the minor-axis (PA=141{\degr}) from a pseudo-slit centered on TH2 (Fig.~\ref{fig7}, middle). The emission along this axis is resolved into the primary bright core following a steep velocity gradient, the extended blue tail on its SE side, and the red component in a bright clump on the NW side.  
The minor-axis velocity curve of the primary component, extracted in the same way as the major-axis velocity curve, is linear within [-0\farcs6,0\farcs6] of TH2 (Fig.~\ref{fig7}) with a best-fit gradient of $+41$ \kms\ arcsec$^{-1}$ from SE to NW. This gradient, however, spans a distance only slightly larger than the spatial resolution of our data. The true velocity curve for for R$\lesssim$5-10 pc might have an even steeper gradient or significantly depart from a linear profile, but such structures would be smeared out in our NIRSPEC cube. Indeed, \citet{rico2006} measure a minor-axis gradient of 110 {\kms}arcsec$^{-1}$ in a high-resolution map ($\sim$0\farcs3 beam) of H92$\alpha$.  

\subsubsection{Substructure of the IR Core and Source S1} \label{subsec:kin3}

A third, distinct kinematic substructure is exhibited by the velocity field of S1 as an 
{ arc of blue-shifted emission} around the east side of the of the IRC {(Fig.~\ref{fig6}, top)}. The S1 blue flow corresponds to the maximum velocity dispersion in the NIRSPEC FoV, reaching $\sigma_{\mbra} \simeq 110$ {\kms} (Fig.~\ref{fig6}, bottom). The gas within the blue knot reaches a maximum LoS velocity shift of $\Delta v= -20$ {\kms} at distance of about $0\farcs8$ from the IRC peak. RRL velocity fields show a ``kink" in the contours near the S1 region likely related to this feature \citep{anantha1996,rico2006}. The blue wing in Figs.~\ref{fig4} and \ref{fig5} contributes the most to the \bra\ line profile at the location of the blue flow, suggesting that these features are physically related. 

\begin{figure*}
\includegraphics[width=0.36\textwidth]{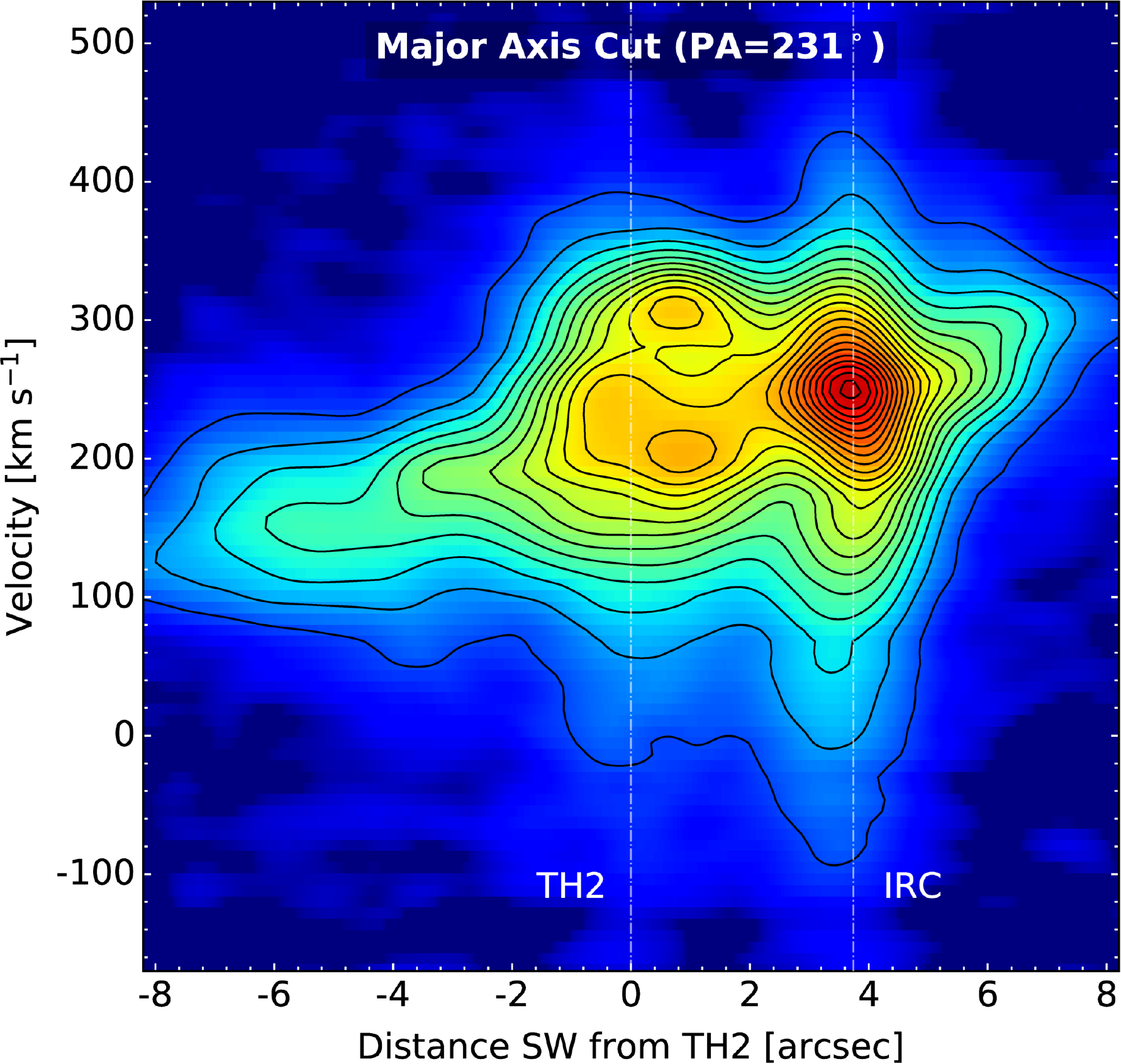}
\includegraphics[width=0.24\textwidth]{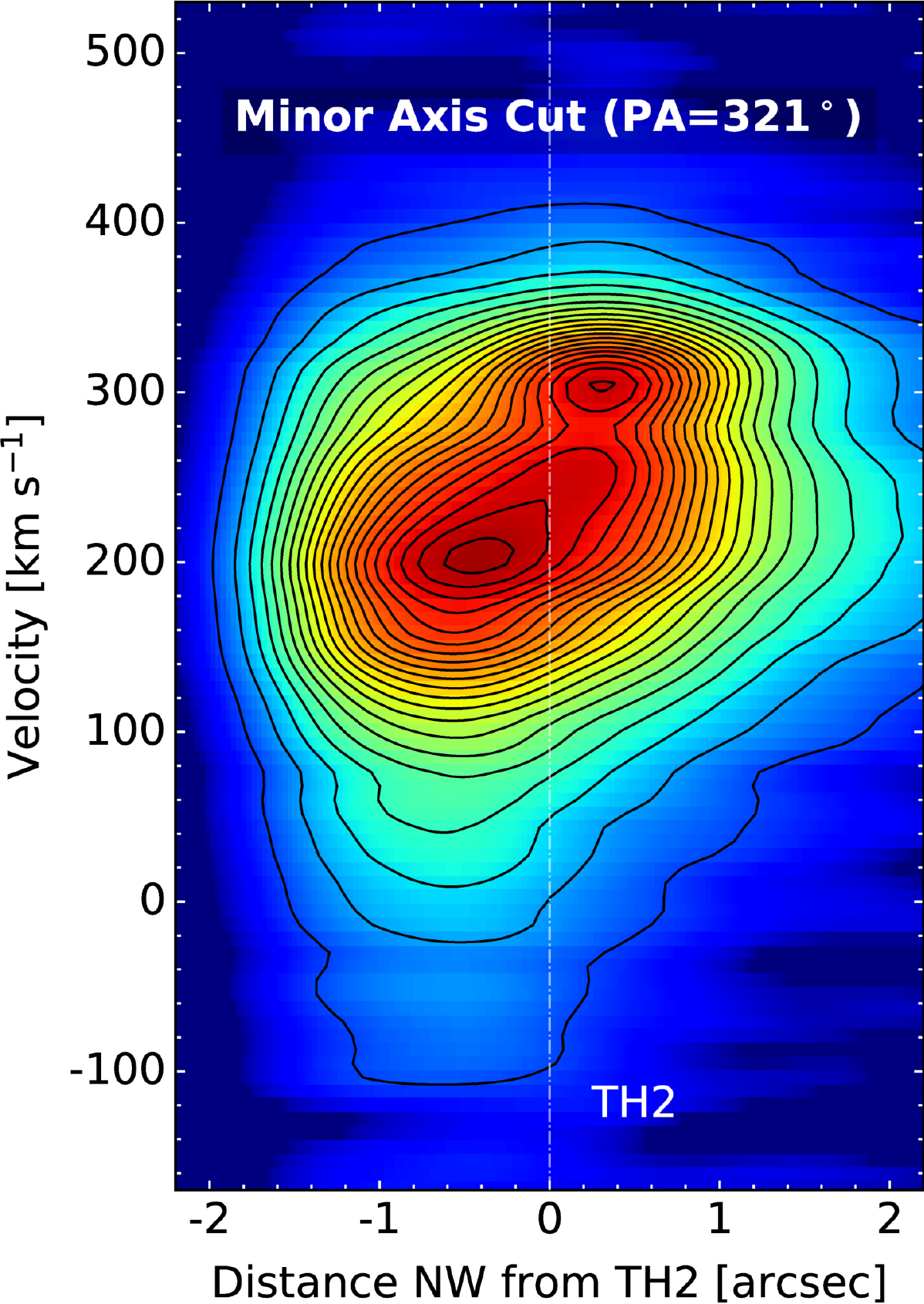}
\hspace{1mm}\raisebox{5ex}{% 
 \includegraphics[width=0.3\textwidth]{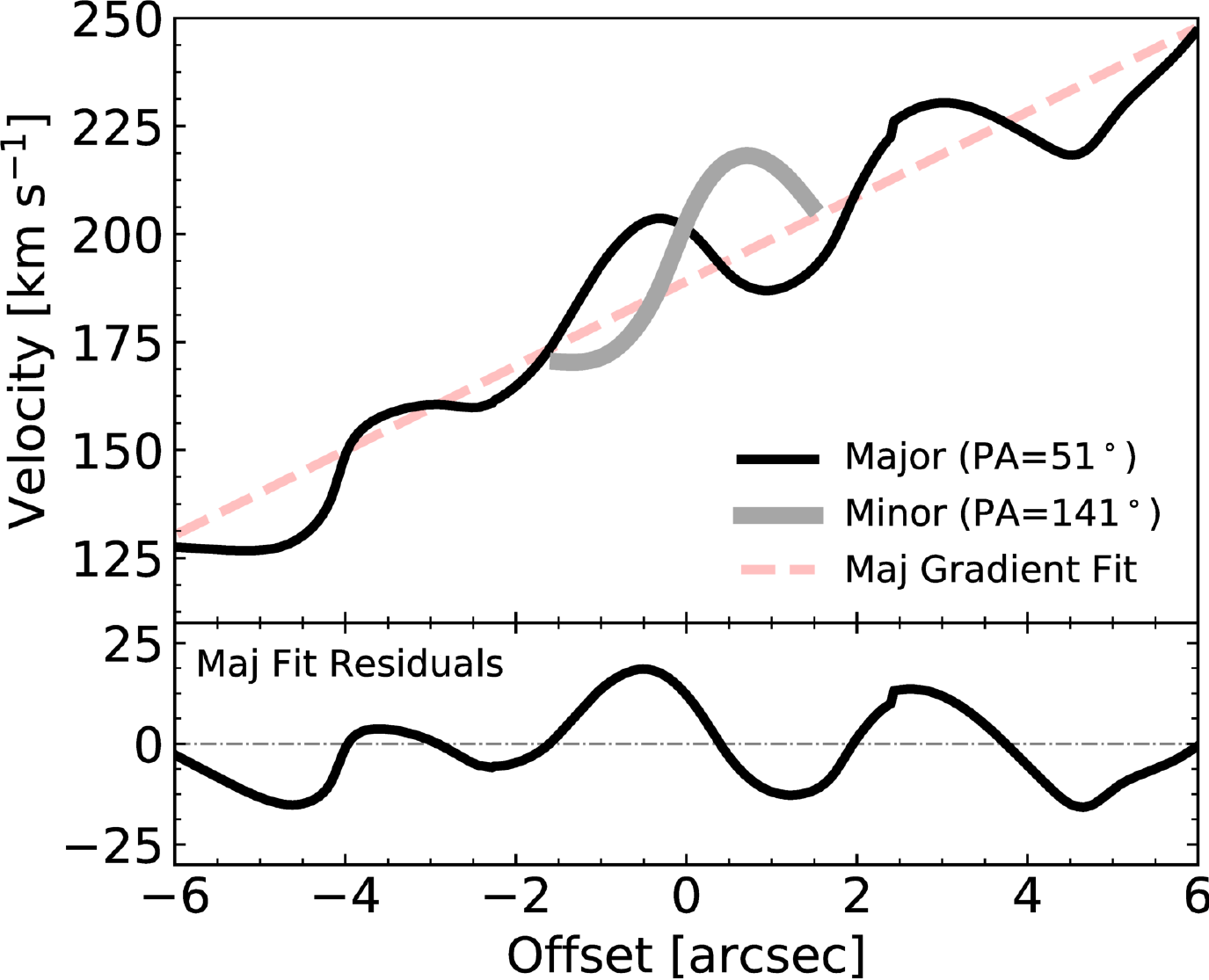}}
\centering
\caption{Position-velocity images along the major (left panel) and minor axes (middle panel), along with velocity profiles extracted from these PV images (right panel). The pseudoslits used to generate the PV images, shown in Fig.~\ref{fig6} as dashed black arrows, are centered on the radio peak TH2, which corresponds to $x=0$ in all panels. The IRC/radio source TH7 is at a position $x\simeq 3\farcs3$ in the major-axis diagram. 
\label{fig7}}
\end{figure*}

The major-axis PV diagram (Fig.~\ref{fig7}, left) detects S1 as a single brightest emission peak with {broad, extended curved wings on both the red and blue sides of the line}. The central peak shows a velocity gradient with opposite orientation to the smooth major-axis gradient, which is exhibited in Fig.~\ref{fig7}, right, as a residual from the linear fit at $x\simeq3\arcsec$, corresponding to a $-$15 {\kms} arcsec$^{-1}$ gradient across the IRC. \citet{gunthardt2015} finds a similar negative gradient of about $-$20 \kms\ arcsec$^{-1}$ for the $\sim$1.5\arcsec\ across the IRC in their H$_2$ rotation curve.
Unfortunately the NIRSPEC slit coverage does not extend more than $\sim$1-1.5\arcsec\ to the SE side of the IRC, such that we cannot determine if the blue flow in Fig.~\ref{fig6} is part of a larger kinematic substructure.

\section{DISCUSSION} \label{sec:discussion}

Kinematic structures in the \bra\ velocity field (Figs.~\ref{fig6} and \ref{fig7}) are linked primarily to the brightest sources sources in the region: N1 {at} the radio core and S1 {at} the IR core. These gas motions reflect the fundamental, complex mechanisms of secular galaxy evolution and the mass evolution of the centers of galaxies.

\subsection{Gaseous Bar Structure Near the Radio Nucleus} 
The strongest kinematic substructure, associated with N1, is distinguished by S-shaped velocity contours. S-shaped contours are characteristic of motion in a barred potential \citep[e.g.,][]{athana1984,athana1999,kormendy2004}. 

Contours in the \bra\ velocity field (Fig.~\ref{fig6}) are oriented at PA$\simeq$45$^{\circ}$, close to the disk major-axis (PA=51{\degr}) and aligned with the nuclear thermal radio continuum knots \citep{turner1983,turner1985,ulvestad1997}. 
This is exactly the orientation of sky-projected inner $x_2$ orbits in the barred potential model from \citet{das2001}, which was used to explain the H92$\alpha$ \citep{anantha1996} and CO velocity fields. In the \citet{das2001} model, the $x_1$ orbits, along the bar's major axis, form the larger NIR bar structure observed at PA=$70^\circ$ \citep{scoville1985,arnaboldi1995,peng1996,iodice2014}. We suggest that gas within the inner ring of $x_2$ orbits (along the NIR bar's minor axis) comprises a secondary nuclear bar that dominates ionized gas velocity structure {in the radio core}.

The candidate nuclear bar in NGC~253 is small. Its length $L_{\mathrm{bar}}\simeq5\arcsec\sim85$ pc (Fig.~\ref{fig6}) is only $\sim1$\% the size of the primary galactic bar in NGC~253, which has length $\sim$7 kpc \citep[e.g.][]{scoville1985,arnaboldi1995,das2001}. The \bra\ velocity profile indicates that is it rapidly rotating. The angular speed of the solid-body disk rotation is constant at $\Omega\simeq590$ {\kms} kpc$^{-1}$. However, the gradient is steeper across the bar region, roughly -20 {\kms} arcsec$^{-1}$ along the major axis within $R\lesssim1\farcs0$, corresponding to an angular speed of $\Omega\sim1200$ {\kms} kpc$^{-1}$ (Fig.~\ref{fig7}). In reality, the secondary bar rotation speed is somewhere in between, $\Omega_{s,\mathrm{bar}}\sim 600$-1200 {\kms} kpc$^{-1}$.  
At $R\sim 50$ pc, the angular speed is $\gtrsim$10$\times$ faster than the primary bar's pattern speed of $\Omega_{p,\mathrm{bar}}\sim 50$ {\kms} kpc$^{-1}$. The result, albeit {uncertain due to inevitable contributions of star cluster feedback to the velocity field, may be compared with the few existing measurements of inner bar rotation speed. Velocity maps produced from H~$\alpha$ observations generally suggest secondary nuclear bars rotate $\gtrsim$3$\times$ faster than primary bars, reaching $\Omega \sim100$-500 {\kms} kpc$^{-1}$ \citep[e.g.,][]{font2014}. Although this is slow compared to the rotation implied by \bra\ for NGC~253, the discrepancy might arise because the H~$\alpha$ emission traces an unobscured, extended gas component, while Br~$\alpha$ emission can originate in obscured gas within a dusty bar}.

{The \bra\ velocity structure presented here is perhaps the strongest evidence yet for a nuclear bar in NGC~253. While tentative evidence for arcing contours has been found in RRL velocity fields, beam smearing and insufficient velocity resolution wash out the characteristic S-shaped bending \citep[e.g.,][]{anantha1996,das2001,rico2006}. Using $K$ band observations, \citet{rosenberg2013} is the only other investigation to unambiguously resolve this structure, to the best of our knowledge. Interestingly, the S-shaped pattern at the radio core was found not only in their H$_2$ velocity field, but also in their {\it stellar} velocity field, strongly suggesting a gravitational origin. Conversely, the IRC's blue flow identified in our {\bra} velocity field is marginally seen in the H$_2$ field but has no counterpart in the stellar field, indicating gas dynamics shaped by star formation. We conclude that the radio core comprises gas and forming massive star clusters, orbiting within the strong gravitational potential of a $\lesssim$100 pc long nuclear bar. Constraining the properties of the bar, such as pattern speed and morphology, will require further observational study.}

\subsection{The Radio Core as the Galactic Center} 

The build-up of massive galactic bulges is largely driven by the formation and feedback of SMBHs, which are ubiquitous in massive galaxies. 
 The symmetry of velocity contours near N1 favors the radio core, close to the brightest radio source TH2, as the location of the galaxy's kinematic center, rather than the IRC. The nature of TH2 is elusive. Its high brightness temperature \citep{turner1985} 
 and lack of associated mid-infrared continuum emission indicates a non-thermal emission mechanism, 
 strongly disfavoring a \hii\ region powered by super star clusters as its origin. Rather, the two leading explanations of TH2 are: an unusually bright supernova remnant (SNR), or an AGN \citep{turner1985,ulvestad1997,mohan2002}. TH2 would be an unusually bright SNR, $\sim$100 times brighter than the Galaxy's brightest SNR, Cas A. Although there is no X-ray or IR counterpart source associated with TH2 \citep{ontiveros2009,muller2010}, the presence of high-$J$ CO, HCN, and dust emission suggests very high extinctions \citep{bolatto2013,meier2015,leroy2018}. If TH2 is indeed a weak AGN, it would be a valuable nearby example of the starburst-AGN interaction.
 
 The \bra\ velocity profile across TH2 yields mass estimates for a potential central SMBH in the radio core (Fig.~\ref{fig7}). The minor-axis velocity gradient,
 roughly 50 \kms\ over 1\farcs5, suggests 
$M_{\mathrm{dyn}}^{\mathrm{TH2}}\sim 1.4 x 10^7$ {\msun}. A similar mass of $M_{\mathrm{dyn}}^{\mathrm{TH2}}\sim 7\times 10^6$ {\msun}, based on a $\sim$110 {\kms} arcsec$^{-1}$ gradient, is derived from the $0\farcs3$-beam H92$\alpha$ map in \citet{rico2006}. Given the measured central stellar velocity dispersion of NGC~253 \citep[$\sigma = 109$ \kms;][]{oliva1995}, the $M$-$\sigma$ relation \citep{combes2019} predicts a SMBH of mass $M_{\mathrm{bh}} \sim 2 \times 10^7$ {\msun}, consistent with the estimate from {\bra}. The predicted sphere of influence has radius $r_h \sim G M_{\mathrm{bh}} / \sigma^2 \sim 8$ pc \citep{merritt2004}. 
The true velocity profile within the black hole's sphere of influence could be characterized by a significantly steeper gradient than is measured after being smeared out by the PSF of our observations. 
Extremely high extinction ($A_V \gtrsim 50$) in the center of NGC~253 \citep[e.g.,][]{leroy2018}, should also significantly affect measurements of the velocity field of IR emission lines and thus estimates of dynamical mass. 

High-resolution observations of gas in the centers of Seyfert galaxies suggest nuclear bars may be intimately tied to AGN phenomena, closely linked to the feeding and growth of SMBHs \citep[e.g.,][]{onishi2015,barth2016,davis2017,davis2018}. Given the bar-like kinematic substructure from our \bra\ measurements, the radio core of NGC~253 near TH2 is an excellent target for high-resolution IR-radio gas spectroscopy.

\subsection{Starburst Feedback: Outflow from the IRC?}

The starburst region of NGC~253 is driving a large-scale galactic wind, with an ionized outflow extending to $\sim$10 kpc \citep[e.g.,][]{strickland2002,weaver2002,westmoquette2011,gunthardt2019} along with molecular outflow from the central $\lesssim 1$kpc \citep[e.g.,][]{sakamoto2006,bolatto2013,walter2017}. The galactic wind should be powered by feedback from SSCs identified in the central starburst region \citep[e.g.,][]{turner1985,watson1996,ulvestad1997,engelbracht1998,forbes2000,ontiveros2009,leroy2018}. 
{Characterization of outflow sources in NGC 253 is key for understanding the link between feedback from its starburst and its galactic wind.}
 
{Source S1 at the IRC exhibits the best evidence for outflow from its line profile and kinematic substructure}. Assuming a mass of $10^6$ {\msun} \citep{leroy2018}, the predicted escape velocity from the IRC is $v_{\mathrm{esc}}(R=20~\mathrm{ pc})\sim 20$ {\kms}. This is significantly smaller than the {\bra} line width of S1, $\sigma_v$ $\sim$50, 130 {\kms} for the primary and broad {\bra} components, respectively (Fig.~\ref{fig4}). A maximum outflow velocity can be estimated with the velocity offset between the broad and primary components and the broad-component line width: $v_{\mathrm{max}} = \Delta v_{b,p} -  \mathrm{FWHM}_{\mathrm{broad}}/2$ \citep[e.g.,][]{veilleux2005,arribas2014,wood2015}. Using average values for S1, $\Delta v_{b,p} \simeq -75$ {\kms} and {FWHM}$_{\mathrm{broad}}\simeq 300$ {\kms}, yields $v_{\mathrm{max}}\sim -225$ {\kms}, consistent with the Br~$\gamma$ outflow \citep{gunthardt2019}. 

{The potential outflow from the IRC is identified as a blue-shifted substructure in the {\bra} velocity field distinct from the nuclear bar pattern (Fig.~\ref{fig6}), and as a negative velocity gradient and with curving red and blue wings in the major-axis PV image (Fig.~\ref{fig7}). These structures suggest de-projected outflow speeds of $\gtrsim 100$ {\kms}, consistent with estimates from the line profile. We conclude that the {\bra} outflow from the IRC forms the base of the $\sim 100-300$ {\kms} H$\alpha$ outflow \citep{westmoquette2011}. Feedback from $\sim$6-8 Myr-old SSC in the IRC appears presently capable of powering the kpc-scale galactic wind.}

Like the \bra\ emission from S1, the emission from N1 has a broad blue component with line width exceeding the local escape velocity, indicating star formation feedback. However, the nuclear bar clearly dominates the velocity structure in the radio core. Compared with the IRC, the radio core exhibits a significantly larger Brackett line equivalent width, $W(\mbra)\sim 250$ {\AA} (Sec.~\ref{subsec:lineprofile}), and extinction \citep[$A_V\sim20-4000$;][]{gunthardt2015,leroy2018}. This suggests a younger stellar population within the radio core formed in a more recent starburst episode. Feedback might be correspondingly weaker in this region if clusters have not yet evolved WR winds and SNe capable of driving significant mass loss. {Alternatively, feedback might be suppressed as newly formed clusters separate from their natal gas within the bar potential due to tidal interactions. NGC~253 demonstrates the complex interplay between nuclear gravitational structure and star formation shaping the evolution of its central bulge.}

\section{SUMMARY} \label{sec:summary}

Using NIRSPEC we have mapped the \bra\ line in the core of NGC~253, the site of a forming galactic center and an intense starburst that is powering a galactic wind on much larger scales. The constructed \bra\ cube, with $1\arcsec$ and 12 {\kms} resolutions, reveals {four} sources associated with complex gas motions influenced by nuclear galactic structure along with feedback from forming massive clusters. The two brightest sources, which we identify as N1 and S1, are associated with the bright radio and IR cores, respectively. Our main findings are as follows:

\begin{enumerate}

\item An underlying gradient of 10 {\kms} arcsec$^{-1}$ across $>10\arcsec$ along the major-axis is identified as the solid-body rotation curve of the inner disk. Residuals of a linear fit appear sinusoidal, with peaks of amplitude $\sim$15 \kms\ offset roughly $0\farcs6\simeq 10$ projected-pc to the NE of each \bra\ sources. The residuals reflect non-circular motions; likely tracing structure of an inner nuclear bar/nuclear spiral.

\item S-shaped contours in the velocity field near N1 provide strong evidence for a nuclear gaseous bar centered on the radio core oriented at PA$\simeq$45{\degr}, 
aligned with {predicted inner $x_2$ orbits of the larger galactic bar} \citep{das2001,paglione2004}. The inner bar has  
radius $R\sim 40-50$ pc and is rotating rapidly, with $\Omega \gtrsim 600$ km s$^{-1}$. 
Intense star formation indicated by candidate SSCs and SNRs suggests that the nuclear bar can induce inflow into the galactic center and fuel star formation.

\item The galactic center, expected to host a SMBH, is near the radio peak TH2 in the radio nucleus rather than in the IRC. 
Based on the minor-axis gradient of 41 \kms\ arcsec$^{-1}$ across TH2, we derive a mass of $ \sim 10^7$ {\msun} across 1\farcs5$\simeq$25 pc, which can be compared with $M\sim 7\times 10^6$ \msun derived previously from subarcsecond resolution
RRL observations \citep{rico2006}. The sphere of influence of this black hole is $\lesssim 8$~pc in radius. 
The values are in rough agreement, given the lower spatial resolution of these 
data.  The mass expected from the M-$\sigma$ relation, $M\sim 10^7$ \msun, is consistent with the measurements. 

\item  The kinematics of source S1 indicate outflow driven by the IRC. Broad emission contributes half of the total line flux here, and has a line width greatly exceeding local escape velocities (FWHM$\sim$300-350 {\kms}). The velocity field shows the outflow as a blue-shifted arc-like substructure at the IRC, distinct from the nuclear bar pattern. The estimated maximum outflow speed is $|v_{\mathrm{max}}|\sim 200-250$ {\kms}, consistent with the H~$\alpha$ outflow \citep[][]{westmoquette2011}. The results provide a plausible link between feedback from star formation in the IRC and the observed large-scale galactic wind.

\end{enumerate}

With a potential nuclear bar and SMBH, the radio core of NGC~253 is an ideal target for high-resolution IR-radio spectroscopic observations. Such measurements will provide insight into the nature of the nonthermal radio source, the presence and influence of a SMBH, and operation of feedback from individual SSCs in the forming galactic center.

\section*{Acknowledgements}

The data presented herein were obtained at the W. M. Keck Observatory, which is operated as a scientific partnership among the California Institute of Technology, the University of California and the National Aeronautics and Space Administration. The Observatory was made possible by the generous financial support of the W. M. Keck Foundation. The authors wish to recognize and acknowledge the very significant cultural role and reverence that the summit of Maunakea has always had within the indigenous Hawaiian community.  We are most fortunate to have the opportunity to conduct observations from this mountain.

%%%%%%%%%%%%%%%%%%%%%%%%%%%%%%%%%%%%%%%%%%%%%%%%%%



\bibliographystyle{mnras}
\tracingmacros=1
\bibliography{ngc253bib} % if your bibtex file is called example.bib

%\begin{thebibliography}{99}
%\end{thebibliography}

%%%%%%%%%%%%%%%%%%%%%%%%%%%%%%%%%%%%%%%%%%%%%%%%%%

%%%%%%%%%%%%%%%%% APPENDICES %%%%%%%%%%%%%%%%%%%%%

%%%%%%%%%%%%%%%%%%%%%%%%%%%%%%%%%%%%%%%%%%%%%%%%%%

% Don't change these lines
\bsp	% typesetting comment
\label{lastpage}
\end{document}